\documentclass[conference]{IEEEtran}
\usepackage{amsmath,amsfonts}
\usepackage{algorithmic}
\usepackage{algorithm}
\usepackage{array}
\usepackage[caption=false,font=normalsize,labelfont=sf,textfont=sf]{subfig}
\usepackage{textcomp}
\usepackage{stfloats}
\usepackage{url}
\usepackage{verbatim}
\usepackage{graphicx}
\usepackage{cite}
\usepackage{booktabs}
\usepackage[hidelinks]{hyperref}
\begin{document}

\title{Configurable p-Neurons Using Modular p-Bits}

\author{Saleh Bunaiyan\textsuperscript{1,2$\dagger$}, Mohammad Alsharif\textsuperscript{3,4$\dagger$}, Abdelrahman S. Abdelrahman\textsuperscript{1}, Hesham ElSawy\textsuperscript{5}, \\ Suraj S. Cheema\textsuperscript{6}, Suhaib A. Fahmy\textsuperscript{4}, Kerem Y. Camsari\textsuperscript{1}, and Feras Al-Dirini\textsuperscript{5,6*}\\ \textsuperscript{1}ECE, UCSB, Santa Barbara, CA, USA, \textsuperscript{2}EE, KFUPM, Dhahran, KSA, \textsuperscript{3}COE, KFUPM, Dhahran, KSA, \\  \textsuperscript{4}CEMSE, KAUST, Thuwal, KSA,  \textsuperscript{5}School of Computing, Queen's University, Kingston, ON, Canada,   \\   \textsuperscript{6}Research Laboratory of Electronics, MIT, Cambridge, MA, USA, \\ $^\dagger$equally contributing authors, *email: {\underline{aldirini@mit.edu}}}

\markboth{Journal of XXXXXXXXX,~Vol.~XX, No.~XX, XXX-XXXX}%
{Shell \MakeLowercase{\textit{et al.}}: A Sample Article Using IEEEtran.cls for IEEE Journals}

\maketitle

\begin{abstract}
Probabilistic bits (p-bits) have recently been employed in neural networks (NNs) as stochastic neurons with sigmoidal probabilistic activation functions. Nonetheless, there remain a wealth of other probabilistic activation functions that are yet to be explored. Here we re-engineer the p-bit by decoupling its stochastic signal path from its input data path, giving rise to a modular p-bit that enables the realization of probabilistic neurons (p-neurons) with a range of configurable probabilistic activation functions, including a probabilistic version of the widely used \textit{Logistic Sigmoid}, \textit{Tanh} and \textit{Rectified Linear Unit (ReLU)} activation functions. We present spintronic (CMOS + sMTJ) designs that show wide and tunable probabilistic ranges of operation. Finally, we experimentally implement digital-CMOS versions on an FPGA, with stochastic unit sharing, and demonstrate an order of magnitude (10x) saving in required hardware resources compared to conventional digital p-bit implementations. 

\end{abstract}


\section{Introduction}
\begin{figure}[!t]
\includegraphics[width=\columnwidth]{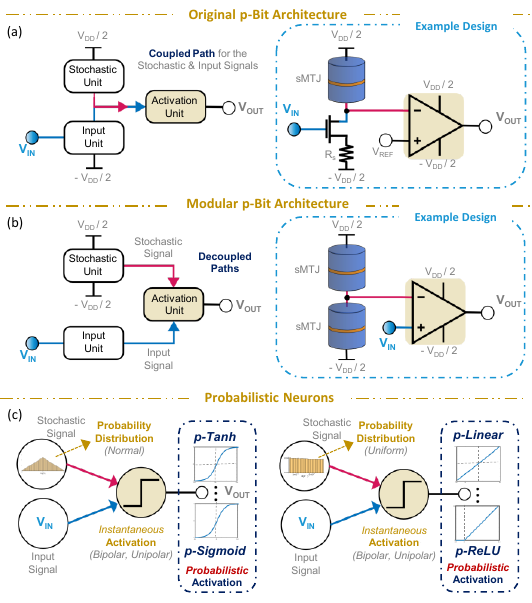}
    \centering
    \caption{P-neurons via decoupled modular p-bits. (a) Original p-bit architecture where the stochastic and input paths are coupled. The original design is shown, where the two paths are coupled at the drain node. (b) Proposed architecture that decouples the stochastic path from the input path, decoupling the design of these two paths. A design example based on a dual sMTJ voltage divider cell is shown. (c) P-neurons with different probabilistic (time-average) activation functions by modular engineering of the stochastic and the activation units.}
    \label{Decoupled_paths}
\end{figure}

\IEEEPARstart{W}{ith} increasing demand for data-intensive computing, unconventional paradigms are emerging~\cite{Neuromorphic_Spintronics}, including probabilistic computing, in which probabilistic-bits (p-bits) based on stochastic magnetic tunnel junctions (sMTJs)~\cite{charge_spin_qbit, p_bit_review, Implementing_pbits, Integer_factorization,situ_aware, Saleh_p-bit,augmented_p_bit,Saleh_p-sensing} are central building blocks. Recently, there has been increasing interest in the realization of neural networks (NNs) employing p-bits as stochastic neurons~\cite{singh2023hardware,liu2023creating}, which have shown high energy-efficiency and low area requirements, compared to deterministic NNs~\cite{singh2024cmos}. The p-bit was first proposed as a modified version of magnetic random access memory (MRAM) technology~\cite{Implementing_pbits}, where the main difference is that the MTJ is made stochastic rather than being deterministic (Fig.~\ref{Decoupled_paths}~(a)).

In this work, we also adopt another inspiration from MRAM technology; the separation between -- or the decoupling of -- the write and read paths~\cite{MRAM_Review, p_bit_big_review}. We propose a novel decoupling approach in the p-bit between the stochastic path and the input path, such that the effect of each of the two paths on the p-bit response is independent of the other. Accordingly, each path can be engineered separately, resulting in a modular p-bit (Fig.~\ref{Decoupled_paths}~(b)) \cite{PatentDecoupledPBit,Patent_tunable_p-bit}. We leverage this approach to customize the response of the p-bit, implementing probabilistic neurons (p-neurons) \cite{PatentPNeurons} with probabilistic versions of widely used activation functions in NNs; namely \textit {Tanh}, \textit {Logistic Sigmoid}, and \textit {Rectified Linear Unit (RELU)}. Transistor-level analog spintronic (CMOS + sMTJ) designs are shown in Fig.~2, while digital CMOS designs are shown in Fig.~\ref{Digital_Neuron}.

\section{Decoupled Architecture for Modularity}
\begin{figure*}[!t]
    \centering
    \includegraphics[width= 7.1 in]{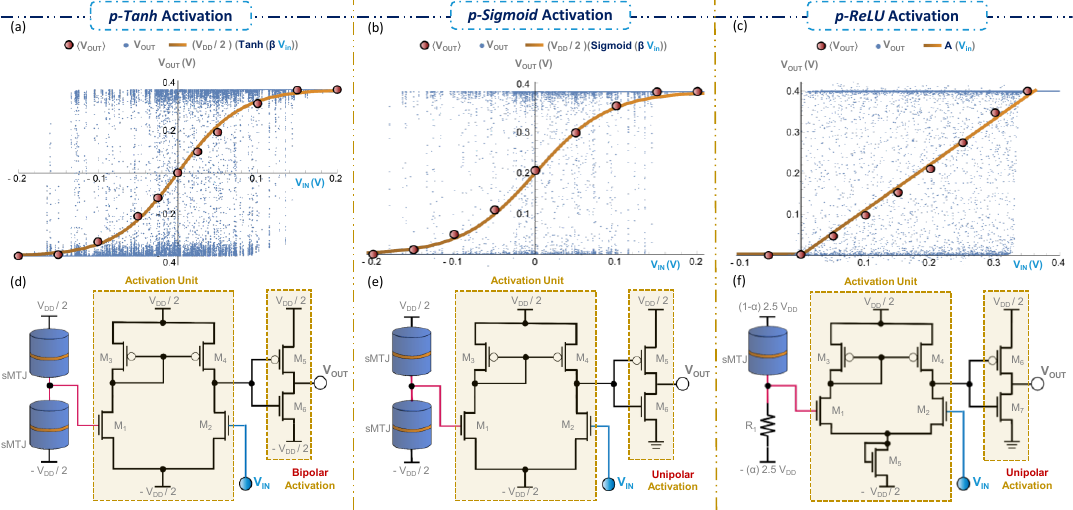}
    \vspace{-3mm}
    \caption{Probabilistic Neurons: spintronic (CMOS + sMTJ) design examples. (a)--(c) response of p-neurons (blue dots: instantaneous response, large orange circles: time-averaged response) with \textit{p-Tanh}, \textit{p-Sigmoid}, and \textit{p-RELU} activation functions, respectively.  Non-bipolar data points are due to the limited slew-rate of the amplifier \cite{Mohammed_2025,GeneralizedOTA,OTAMWSCAS,ScalableOTAs}. (d) and (e) transistor-level circuit designs for implementing p-neurons with \textit{p-Tanh} and \textit{p-Sigmoid} activation functions, respectively. The stochastic unit is a dual sMTJ voltage divider (2M cell). For \textit{p-Sigmoid}, the activation unit was re-engineered to obtain and optimize unipolar activation of the p-neuron through the reduction of the $W/L$ of $M_6$ by a factor of three. (f) transistor-level circuit design for implementing a p-neuron with a \textit{p-RELU} activation function, using a single sMTJ + single resistor (1M1R) cell as the stochastic unit, with $R_1$ = $0.35/G_0$ and $\alpha = 0.155$. For all designs $V_{\mathit{DD}}$ = 0.8.}
    \label{Activation}
\end{figure*}

\begin{figure}[!b]
\includegraphics[width=\columnwidth]{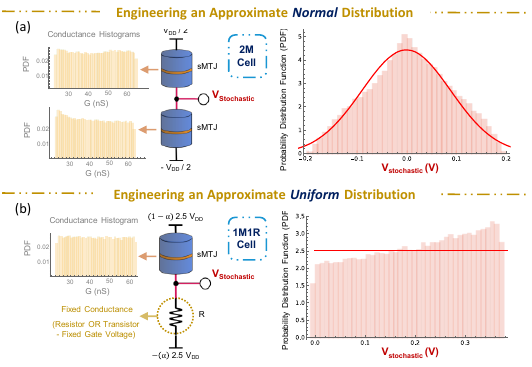}
    \centering
    \caption{Engineering the stochastic unit: (a) A dual sMTJ stochastic unit - 2M cell (left), consisting of two sMTJs in series (each with a uniform conductance distribution). The probability distribution of $V_{Stochastic}$ of the 2M cell that is approximately normal (right). (b) A stochastic unit that consists of one sMTJ and one fixed resistor connected in series - 1M1R cell (left), with $R_1$ = $0.35/G_0$ and $\alpha = 0.155$. The probability distribution of $V_{Stochastic}$ of the 1M1R cell that is approximately uniform (right).}
    \label{Stochasticity}
\end{figure}

In the original \textit{``Coupled Architecture''} of the p-bit changing the input directly alters the stochastic signal generated by the stochastic unit. A common example is the original p-bit design~\cite{Implementing_pbits}, shown in Fig.~\ref{Decoupled_paths}~(a), where the input voltage $V_{\mathit{IN}}$ directly affects the stochastic response at the drain node~\cite{Implementing_pbits, Saleh_p-bit}. On the other hand, in our \textit{``Decoupled Architecture''}, the effect of the stochastic unit response is decoupled from the input signal (Fig.~\ref{Decoupled_paths}~(b)). A dual sMTJ voltage divider (2M cell) is shown in Fig.~\ref{Decoupled_paths}~(b) as an example design of the stochastic unit. Decoupled designs can be utilized to generalize the p-neuron response to any probabilistic response with proper engineering of the stochastic unit and the instantaneous activation unit, as shown in Fig.~\ref{Stochasticity}~(a)--(b) and Fig.~\ref{Activation}. In this work, we exploit the modularity of the decoupled architecture, showing that not only can we implement spintronic-based modular p-neuron designs, but digital CMOS versions also, which we experimentally realize using a Field-Programmable Gate Array (FPGA), with customizable and tunable probabilistic activation functions, as well as shared stochastic units.

\begin{figure*}[!t]
\includegraphics[width= 7.1 in]{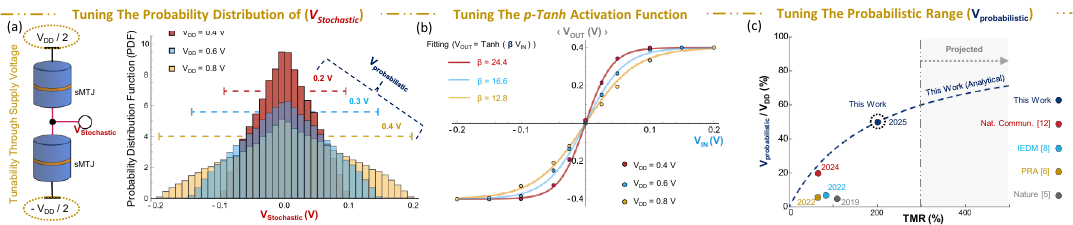}
    \centering
    \vspace{-3mm}
    \caption{Tunability of the p-neuron's probabilistic activation function. (a) For a stochastic unit of two sMTJs in series (2M cell), the characteristics of the $V_{Stochastic}$ probability distribution (mean and variance) can be tuned by $V_{DD}$. (b) Tuning the p-neuron's probabilistic range of the \textit{p-Tanh} activation function. The \textit{p-Tanh} tunability is controlled by a scaling factor $\beta$. (c) Analytical limit that describes the tunability of the probabilistic range ($V_{probabilistic})$ as a function of the sMTJ tunneling magnetoresistance (TMR). The other data points refer to other experimental realizations of p-bits.}
    \label{Tunable}
\end{figure*}

\section{Spintronic Designs (CMOS + Stochastic MTJs) }
The first decoupled design example employs a stochastic unit of two identical sMTJs connected in series in a 2M cell (Fig.~\ref{Stochasticity}~(a)), which we simulate in HSPICE using the stochastic Landau-Lifshitz-Gilbert (sLLG) equation~\cite{butler2012switching}. We consider an sMTJ with perpendicular magnetic anisotropy (PMA) and no effective field ($\Delta_B \approx \vec H_{eff} \approx 0$). The magnetization $\mathbf{\hat{m}}$ of such an sMTJ was theoretically proven to have uniform distribution over all directions~\cite{Quantitative_Evaluation}. Hence, the sMTJ conductance $G(t)$ will also be uniform. We assume a free layer diameter of 22 nm, a polarization of 0.7, and a low $G_0$ to minimize spin-transfer torque (STT); preserving the uniform randomness of $G(t)$, while keeping other simulation parameters as in previous work~\cite{Implementing_pbits}. Then we couple this stochastic unit to a differential amplifier followed by pull up (PU) and pull down (PD) networks, constructed using an inverter (Fig.~\ref{Activation}~(d)). All CMOS transistors are simulated using the HP 14-nm FinFET predictive technology model (PTM)~\cite{PTM}. The time-averaged response of this design implements a probabilistic Tanh (\textit{p-Tanh}) activation function as shown in Fig.~\ref{Activation}~(a), where the supply voltage swing is bipolar. Note that this realization eliminates the requirement for matching the sMTJ with the transistor. By fixing the stochastic branch and engineering the activation unit, we can adjust the activation function to be a probabilistic sigmoid (\textit{p-Sigmoid}), as in Fig.~\ref{Activation}~(b) and (e). 

Another design example is implemented by replacing one of the sMTJs with a fixed resistance in the stochastic unit, resulting in a 1 sMTJ + 1 resistor (1M1R cell) stochastic unit. In this example a probabilistic ReLU (\textit{p-ReLU}) activation function is realized (see Fig.~\ref{Activation}~(f)), which is achieved by adjusting the voltage $V_{\mathit{Stochastic}}$ between $R_1$ and the sMTJ to have a uniform distribution (Fig.~\ref{Stochasticity}~(b)). The uniform random conductance of the sMTJ does not result in an ideal uniform distribution at $V_{\mathit{Stochastic}}$ in the 1M1R cell, however, it can be adjusted to become near-uniform by choosing a suitable value for $R_1$ and extending the supply voltage of the stochastic unit as shown in Fig.~\ref{Stochasticity}~(b). Other approaches and technologies used in encryption \cite{AFMTRNG,koh2025closed,PatentAdaptiveTRNG,adaptiveTRNGArXiv,SNWMemristor,RSQMemristorScaling} and sensing \cite{PatentPSensing,Sensor_Journal,MWSCAS,ACCESS}, can also provide such a distribution. The time-averaged response of the \textit{p-RELU} activation function is shown in Fig.~\ref{Activation}~(c).

\section{Probabilistic Range Tunability}
Using the supply voltage $V_{\mathit{DD}}$ across the two sMTJs, we can shrink or stretch the range of input voltages to which the p-neuron has a stochastic response ($V_{\mathit{Stochastic}}$), as shown in Fig.~\ref{Tunable}~(a). We define this range of input voltages here as the probabilistic range of the p-neuron ($V_{\mathit{probabilistic}}$). Hence, using $V_{\mathit{DD}}$, we can tune this probabilistic range of operation of the p-neuron, as shown in Fig.~\ref{Tunable}~(b). The value of $V_{\mathit{probabilistic}}$ relative to the overall supply voltage, obtained using the 2M cell stochastic unit, can be described as a function of the sMTJ's tunneling magnetoresistance (TMR):
\begin{equation}
V_{\mathit{probabilistic}} / V_{\mathit{DD}} = {\mathit{TMR}}/({2+\mathit{TMR}}) 
\label{eq: Theory}
\end{equation}
where $\mathit{TMR} = (R_{\mathit{AP}} - R_P)/R_P$, $R_{\mathit{AP}}$ and $R_P$ are antiparallel and parallel resistances of the sMTJ, respectively. This analytical limit of $V_{\mathit{probabilistic}}/V_{\mathit{DD}}$ is plotted in Fig.~\ref{Tunable}~(c), showing that, at high TMR (around 300 $\%$), a probabilistic range of around 60 $\%$ of $V_{\mathit{DD}}$ can be attainable. Although higher values of TMR were reported for stable MTJs~\cite{600TMR_2,600TMR_1}, sMTJs are still at early stages and their limits are yet to be discovered. Nonetheless, the plot shows an enhancement in $V_{\mathit{probabilistic}}$ for the modular p-neuron, across a range of TMR values, compared to previous designs reported in the literature.

\begin{figure*}[!t]
\includegraphics[width= 6.4 in]{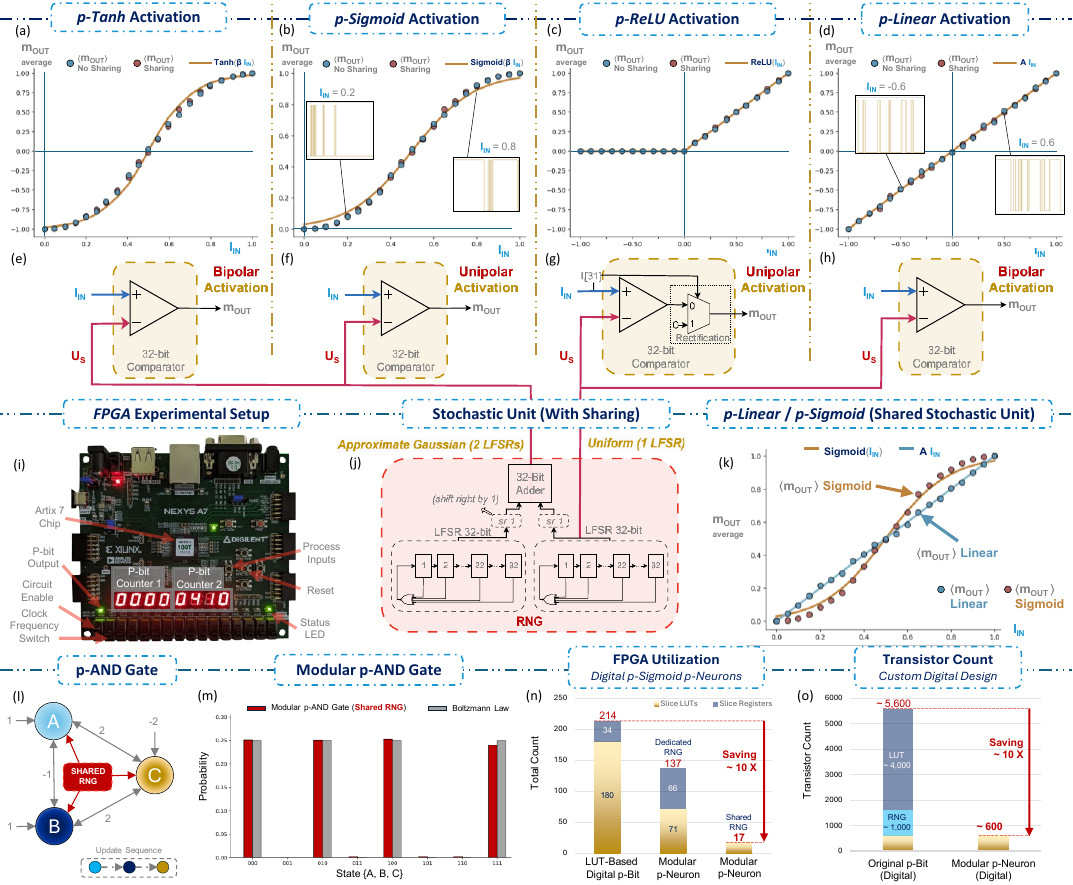}
    \centering
    \vspace{-3mm}
    \caption{Digital p-neurons and stochastic unit sharing experiments. (a)--(d) Time-averaged response of digital p-neurons with (a) \textit{p-Tanh} (b) \textit{p-Sigmoid} (c) \textit{p-ReLU} and (d) \textit{p-Linear} activation functions, implemented on an FPGA. (e)--(h) The p-neuron digital designs employing 32-bit comparators, which compare the input string $I_{\mathit{IN}}$ with the random number generated by the stochastic unit. 
    (i)~Experimental setup for the FPGA implementation. 
    (j)~Shared stochastic unit. (k)~The result of an FPGA experiment where one \textit{p-Sigmoid} p-neuron and one \textit{p-Linear} p-neuron receive their random numbers, with different probability distributions, from a single shared stochastic unit. (l)~A modular probabilistic AND (p-AND) gate with 3 p-neurons that share a single RNG as their stochastic unit. The interconnections between the p-neurons are shown with weights on bidirectional arrows and biases on unidirectional arrows. (m)~Histogram showing the probability distribution of visiting the correct p-AND gate states, and the expected Boltzmann distribution. (n)~Savings in FPGA hardware resource utilization. (o)~Comparison of the estimated transistor count in custom digital designs of modular \textit{p-Sigmoid} p-neurons with original FPGA p-bits \cite{singh2024cmos}.}
    \label{Digital_Neuron}
\end{figure*}

\section{Digital CMOS FPGA Implementations}
To demonstrate the generality of our approach, we reproduce the probabilistic activation functions obtained earlier using digital CMOS implementations on an FPGA, employing linear-feedback shift registers (LFSRs) for the stochastic units. Since an LFSR naturally generates a uniform (pseudo) random variable, we can approximate the response of the \textit{p-Tanh} and \textit{p-Sigmoid} activation functions by using a stochastic unit that employs two 32-bit LFSRs and sums their output using a 32-bit adder. This addition of two uniform distributions constructs an Irwin–Hall distribution that approximates the Gaussian/normal distribution needed for these probabilistic activation functions. The generated stochastic sum (usually referred to as the ``random number'' in digital systems) is compared to the digital input string ($I_{\mathit{IN}}$), represented by a 32-bit fixed point unsigned binary representation, using a 32-bit digital comparator. On the other hand, for the \textit{p-RELU} and probabilistic linear (\textit{p-Linear}) p-neurons, only a single LFSR is required to construct the stochastic unit, as it generates the needed random number with a uniform distribution. In both cases, 2's complement binary representation is used.

The digital designs of each of these p-neurons are shown in Fig.~\ref{Digital_Neuron}~(e)--(h), building upon previous work~\cite{pbits_fpga}. However, all of these modular p-neuron designs, contrary to earlier designs~\cite{pbits_fpga}, do not require a lookup table (LUT) to achieve the required activation function; instead, the input is directly compared to the output of the stochastic unit. It is the stochastic signal's probability distribution that controls and customizes the activation function, minimizing FPGA hardware resource requirements (Fig. \ref{Digital_Neuron}(n)). For the \textit{p-RELU} p-neuron, a rectification unit is needed within the activation unit, as shown in Fig.~\ref{Digital_Neuron}(g), which is a multiplexer that is enabled by the most significant bit of the input $I_{\mathit{IN}}$ (Fig.~\ref{Digital_Neuron}(g)).

\section{Stochastic Unit Sharing}
The experimental results shown in Fig.~\ref{Digital_Neuron}~(a)--(d) not only reproduce the desired activation functions, but also demonstrate how -- enabled by their modular design -- multiple p-neurons can share the same stochastic unit, yet still generate different activation functions (Fig.~\ref{Digital_Neuron}). This is achieved by independently engineering each p-neuron's instantaneous activation unit, as shown in Fig.~\ref{Digital_Neuron}~(e)--(h). This sharing of LFSRs further minimizes FPGA hardware resource requirements, as shown in Fig.~\ref{Digital_Neuron}~(n), reaching beyond an order of magnitude compared to conventional FPGA p-bits. Moreover, in custom digital p-neuron designs, leveraging stochastic unit sharing and the fact that no LUT is needed results in the fact that each additional digital p-neuron requires almost one order of magnitude less transistors, when compared to conventional LUT-based digital p-bit implementations \cite{singh2024cmos}, as shown in Fig.~\ref{Digital_Neuron}(o). Stochastic unit sharing can also be implemented between multiple p-neurons with different classes of activation functions, where each neuron requires a different stochastic unit as in Fig.~\ref{Digital_Neuron}(k). Power savings are also expected; but require future analysis.

An example 3 p-neuron circuit is shown in Fig.~\ref{Digital_Neuron}(l), implementing a probabilistic AND (p-AND) gate that can be operated in both forward and reverse modes. All p-neurons have \textit{p-Sigmoid} activation functions and are configured in a fully connected Boltzmann machine configuration. All p-neurons share the same stochastic unit. The probability distribution of visiting states for the 3 p-neuron network, shown in Fig.~\ref{Digital_Neuron}(m), confirms the network's functionality as a p-AND gate, and matches the expected Boltzmann distribution. 

\section{Conclusion}
We presented spintronic (CMOS + sMTJ) and digital p-neuron designs based on modular p-bits, exhibiting a range of configurable probabilistic activation functions that are highly tunable, and a modular ability of stochastic unit sharing. The digital p-neuron designs do not require LUTs, providing further savings in hardware resources. Our work paves the way toward accessible and scalable hardware implementations of Probabilistic Neural Networks using modular p-neurons. 

\IEEEtriggeratref{19}
\bibliographystyle{IEEEtran}
\bibliography{library}

@ARTICLE{Sensor_Journal,  author={Bunaiyan, Saleh and Al-Dirini, Feras},  journal={IEEE Sensors Journal},   title={Neuro-Inspired Autonomous Data Acquisition for Energy-Constrained {IoT} Sensors},   year={2022}, month={Oct.},  volume={22},  number={20},  pages={19466-19479},  doi={10.1109/JSEN.2022.3200627}}

@ARTICLE{ACCESS,
  author={Attia, Hussein and Gaya, Sagiru and Alamoudi, Abdullah and M. Alshehri, Fahad and Al-Suhaimi, Abdulrahman and Alsulaim, Nawaf and M. Al Naser, Ahmad and Aghyad Jamal Eddin, Mohamad and M. Alqahtani, Abdullah and Prieto Rojas, Jhonathan and Al-Dharrab, Suhail and Al-Dirini, Feras},
  journal={IEEE Access}, 
  title={Wireless Geophone Sensing System for Real-Time Seismic Data Acquisition}, 
  year={2020},
  volume={8},
  number={},
  month={Apr.},
  pages={81116-81128},
  doi={10.1109/ACCESS.2020.2989280}}

@INPROCEEDINGS{MWSCAS,
  author = {Bunaiyan, Saleh and Al{-}Dirini, Feras},
  booktitle={2021 IEEE International Midwest Symposium on Circuits and Systems (MWSCAS)}, 
  title={Real-Time Analog Event-Detection for Event-Based Synchronous Sampling of Sparse Sensor Signals}, 
  year={2021},
  volume={},
  number={},
  pages={1053-1057},
  doi={10.1109/MWSCAS47672.2021.9531687}}

@misc{PatentDecoupledPBit,
  author       = {Al-Dirini, Feras Mohamad Ameer and Bunaiyan, Saleh Ahmad S.},
  title        = {P-bit generator and methods for tuning a P-bit generator having decoupled stochastic and control paths},
  howpublished = {U.S.\ Patent Application\ Number\ 18\,949\,434},
  year         = {2025},
  note         = {Filed 15 Nov 2024},
  url          = {https://patents.google.com/patent/US20250167774A1/en}
}

@misc{PatentPSensing,
  author       = {Al-Dirini, Feras Mohamad Ameer and Bunaiyan, Saleh Ahmad S.},
  title        = {Probabilistic autonomous data acquisition using stochastic {MTJ} based p-bits},
  howpublished = {U.S.\ Patent Application\ Number\ 18\,417\,677},
  year         = {2025},
  note         = {Filed 19 Jan 2024},
  url          = {https://patents.google.com/patent/US20250238290A1/en}
}

@misc{PatentPNeurons,
  author       = {Al-Dirini, Feras Mohamad Ameer and Bunaiyan, Saleh Ahmad S. and Alsharif, Mohammad Sharaf F.},
  title        = {Modular Probabilistic Bit for Implementing Stochastic Neurons with Configurable Activation Functions},
  howpublished = {U.S.\ Patent Application\ Number\ 19\,260\,798},
  year         = {2026},
  note         = {Filed 07 Jul 2025},
  url          = {}
}

@misc{PatentAdaptiveTRNG,
  author       = {Al-Dirini, Feras Mohamad Ameer and Zahoor, Furqan and Albulushi, Ibrahim Abdulrahman},
  title        = {Adaptive Random Number Generator for Cryptographic Application},
  howpublished = {U.S.\ Patent Application\ Number\ 19\,170\,745},
  year         = {2026},
  note         = {Filed 04 Apr 2025},
  url          = {}
}

@INPROCEEDINGS{SNWMemristor,
  author={Abdelrahman, Abdelrahman S. and ElSawy, Hesham and Lanza, Mario and Akinwande, Deji and Al-Dirini, Feras},
  booktitle={2023 Silicon Nanoelectronics Workshop (SNW)}, 
  title={Scalability of {h-BN} Based Memristors: Yield and Variability Considerations}, 
  year={2023},
  volume={},
  number={},
  pages={109-110},
  doi={10.23919/SNW57900.2023.10183973}}

@misc{RSQMemristorScaling,
  author = {Abdelrahman, Abdelrahman and ElSawy, Hesham and Yuan, Yue and Cheema, Suraj and Akinwande, Deji and Lanza, Mario and Al-Dirini, Feras},
  title  = {Defect-Aware Extreme Device Scaling Limits of {2D} Memristive Technologies},
  year   = {2025},
  note   = {04 August 2025, Preprint (Version 1) available at Research Square},
  doi    = {10.21203/rs.3.rs-7265912/v1},
  url    = {https://doi.org/10.21203/rs.3.rs-7265912/v1}
}

@misc{adaptiveTRNGArXiv,
      title={Adaptive Variation-Resilient Random Number Generator for Embedded Encryption}, 
      author={Furqan Zahoor and Ibrahim A. Albulushi and Saleh Bunaiyan and Anupam Chattopadhyay and Hesham ElSawy and Feras Al-Dirini},
      year={2025},
      eprint={2507.05523},
      archivePrefix={arXiv},
      primaryClass={cs.ET},
      url={https://arxiv.org/abs/2507.05523}, 
}

@ARTICLE{GeneralizedOTA,
  author={Mohammed, Mahmood A. and Roberts, Gordon W.},
  journal={IEEE Transactions on Circuits and Systems I: Regular Papers}, 
  title={Generalized Relationship Between Frequency Response and Settling Time of CMOS OTAs: Toward Many-Stage Design}, 
  year={2021},
  volume={68},
  number={12},
  pages={4993-5006},
  doi={10.1109/TCSI.2021.3110106}}

@INPROCEEDINGS{OTAMWSCAS,
  author={Mohammed, Mahmood A. and Emara, Ahmed S. and Roberts, Gordon W.},
  booktitle={2024 IEEE 67th International Midwest Symposium on Circuits and Systems (MWSCAS)}, 
  title={Slew-Rate Analysis of Scalable Multi-Stage CMOS Operational Transconductance Amplifiers}, 
  year={2024},
  volume={},
  number={},
  pages={15-19},
  doi={10.1109/MWSCAS60917.2024.10658762}}

@ARTICLE{ScalableOTAs,
  author={Mohammed, Mahmood A. and Roberts, Gordon W.},
  journal={IEEE Transactions on Circuits and Systems I: Regular Papers}, 
  title={Scalable Multi-Stage CMOS OTAs With a Wide CL-Drivability Range Using Low-Frequency Zeros}, 
  year={2023},
  volume={70},
  number={1},
  pages={74-87},
  doi={10.1109/TCSI.2022.3216201}}

@article{koh2025closed,
  title={Closed Loop Superparamagnetic Tunnel Junctions for Reliable True Randomness and Generative Artificial Intelligence},
  author={Koh, Dooyong and Wang, Qiuyuan and McGoldrick, Brooke C and Chou, Chung-Tao and Liu, Luqiao and Baldo, Marc A},
  journal={Nano Letters},
  volume={25},
  number={10},
  pages={3799--3806},
  year={2025},
  publisher={ACS Publications},

doi={10.1021/acs.nanolett.4c05728}
}

@article{AFMTRNG,
  title={Unraveling Origin of Stochasticity in Multi-Filamentary Memristor},
  author={Soh, Keunho and Koo, Seunghoe and Yoon, Byoungjin and Kim, Ji Eun and Chun, Suk Yeop and Hwang, Su In and Jung, Junki and Jang, Ho Won and Hur, Sunghoon and Kim, Kyeongtae and others},
  journal={Advanced Functional Materials},
  pages={e27482},
  year={2026},
  publisher={Wiley Online Library}
}

@ARTICLE{butler2012switching,
  author={Butler, William H and Mewes, Tim and Mewes, Claudia KA and Visscher, PB and Rippard, William H and Russek, Stephen E and Heindl, Ranko},
  journal={IEEE Transactions on Magnetics}, 
  title={Switching Distributions for Perpendicular Spin-Torque Devices Within the Macrospin Approximation},
  month = {Dec.},
  year={2012},
  volume={48},
  number={12},
  pages={4684-4700},
  doi={10.1109/TMAG.2012.2209122}}

@misc{PTM,
  author = {{Predictive Technology Model (PTM)}},
  title = {Predictive Technology Model},
  howpublished = {\url{http://ptm.asu.edu/}},
  note = {Accessed: 2017}
}

@article{liu2023creating,
  title={Creating stochastic neural networks with the help of probabilistic bits},
  author={Liu, Samuel and Incorvia, Jean Anne C},
  journal={Nature Electronics},
  volume={6},
  number={12},
  pages={935--936},
  year={2023},
  month={Dec},
  publisher={Nature Publishing Group UK London},
  doi={10.1038/s41928-023-01088-7},

}

@INPROCEEDINGS{singh2023hardware,
  author={Singh, Nihal Sanjay and Niazi, Shaila and Chowdhury, Shuvro and Selcuk, Kemal and Kaneko, Haruna and Kobayashi, Keito and Kanai, Shun and Ohno, Hideo and Fukami, Shunsuke and Camsari, Kerem Y},
  booktitle={Proceedings of the IEEE International Electron Devices Meeting (IEDM)}, 
  title={Hardware Demonstration of Feedforward Stochastic Neural Networks with Fast {MTJ}-based p-bits}, 
  year={2023},
  volume={},
  number={},
  pages={1-4},
  month={Dec.},
  doi={10.1109/IEDM45741.2023.10413686}}

@ARTICLE{Implementing_pbits,  author={Camsari, Kerem Yunus and Salahuddin, Sayeef and Datta, Supriyo}, 
 journal={IEEE Electron Device Letters},   title={Implementing p-bits With Embedded {MTJ}},  month={Oct.}, year={2017},  volume={38},  number={12},  pages={1767-1770}, 
 doi={10.1109/LED.2017.2768321}}

@article{Integer_factorization,
author = {Borders, William and Pervaiz, Ahmed and Fukami, Shunsuke and Camsari, Kerem and Ohno, Hideo and Datta, Supriyo},
year = {2019},
month = {Sep.},
pages = {390-393},
title = {Integer factorization using stochastic magnetic tunnel junctions},
volume = {573},
journal = {Nature},
doi = {10.1038/s41586-019-1557-9}
}

@article{situ_aware,
  title = {Hardware-Aware In Situ Learning Based on Stochastic Magnetic Tunnel Junctions},
  author = {Kaiser, Jan and Borders, William A. and Camsari, Kerem Y. and Fukami, Shunsuke and Ohno, Hideo and Datta, Supriyo},
  journal = {Physical Review Applied},
  volume = {17},
  issue = {1},
  pages = {014016},
  numpages = {12},
  year = {2022},
  month = {Jan.},
  publisher = {American Physical Society},
  doi = {10.1103/PhysRevApplied.17.014016},
}

@article{Quantitative_Evaluation,
  title = {Quantitative Evaluation of Hardware Binary Stochastic Neurons},
  author = {Hassan, Orchi and Datta, Supriyo and Camsari, Kerem Y.},
  journal = {Physical Review Applied},
  volume = {15},
  issue = {6},
  pages = {064046},
  numpages = {13},
  year = {2021},
  month = {Jun.},
  publisher = {American Physical Society},
}

@ARTICLE{p_bit_review,
  author={Chowdhury, Shuvro and Grimaldi, Andrea and Aadit, Navid Anjum and Niazi, Shaila and Mohseni, Masoud and Kanai, Shun and Ohno, Hideo and Fukami, Shunsuke and Theogarajan, Luke and Finocchio, Giovanni and Datta, Supriyo and Camsari, Kerem Yunus},
  journal={IEEE Journal on Exploratory Solid-State Computational Devices and Circuits}, 
  title={A Full-Stack View of Probabilistic Computing With p-Bits: Devices, Architectures, and Algorithms}, 
  year={2023},
  volume={9},
  number={1},
  pages={1-11},
  month={Mar.}
 }

@article{pbits_fpga,
  author={Pervaiz, Ahmed Zeeshan and Sutton, Brian M. and Ghantasala, Lakshmi Anirudh and Camsari, Kerem Y.},
  journal={IEEE Transactions on Neural Networks and Learning Systems}, 
  title={Weighted  $p$ -Bits for FPGA Implementation of Probabilistic Circuits}, 
  month={Jun.},
  year={2019},
  volume={30},
  number={6},
  pages={1920-1926},
  doi={10.1109/TNNLS.2018.2874565}}

@ARTICLE{charge_spin_qbit,
  author={Camsari, Kerem Y and Debashis, Punyashloka and Ostwal, Vaibhav and Pervaiz, Ahmed Zeeshan and Shen, Tingting and Chen, Zhihong and Datta, Supriyo and Appenzeller, Joerg},
  journal={Proceedings of the IEEE}, 
  title={From Charge to Spin and Spin to Charge: Stochastic Magnets for Probabilistic Switching}, 
  year={2020},
  volume={108},
  number={8},
  month={Feb.},
  pages={1322-1337},
  doi={10.1109/JPROC.2020.2966925}}

@INPROCEEDINGS{augmented_p_bit,
  author={Grimaldi, Andrea and Selcuk, Kemal and Aadit, Navid Anjum and Kobayashi, Keito and Cao, Qixuan and Chowdhury, Shuvro and Finocchio, Giovanni and Kanai, Shun and Ohno, Hideo and Fukami, Shunsuke and Camsari, Kerem Yunus},
  booktitle={Proceedings of the IEEE International Electron Devices Meeting (IEDM)}, 
  title={Experimental evaluation of simulated quantum annealing with {MTJ}-augmented p-bits}, 
  year={2022},
  volume={},
  number={},
  month={Dec.},
  pages={22.4.1-22.4.4},
  doi={10.1109/IEDM45625.2022.10019530}}

@article{p_bit_big_review,
  title = {Roadmap for unconventional computing with nanotechnology},
  volume = {8},
  ISSN = {2399-1984},
  DOI = {10.1088/2399-1984/ad299a},
  number = {1},
  journal = {Nano Futures},
  publisher = {IOP Publishing},
  author = {Finocchio, Giovanni and Incorvia, Jean Anne C and Friedman, Joseph S and Yang, Qu and Giordano, Anna and Grollier, Julie and Yang, Hyunsoo and Ciubotaru, Florin and Chumak, Andrii V and Naeemi, Azad J and others},
  year = {2024},
  month = {Mar.},
  pages = {012001}
}

@article{MRAM_Review,
title = {Spintronics based random access memory: a review},
journal = {Materials Today},
volume = {20},
number = {9},
pages = {530-548},
year = {2017},
month={Nov.},
issn = {1369-7021},
author = {Sabpreet Bhatti and Rachid Sbiaa and Atsufumi Hirohata and Hideo Ohno and Shunsuke Fukami and S.N. Piramanayagam}}

@article{singh2024cmos,
  title = {{CMOS} plus stochastic nanomagnets enabling heterogeneous computers for probabilistic inference and learning},
  volume = {15},
  ISSN = {2041-1723},
  DOI = {10.1038/s41467-024-46645-6},
  number = {1},
  pages = {2685},
  journal = {Nature Communications},
  publisher = {Springer Science and Business Media LLC},
  author = {Singh, Nihal Sanjay and Kobayashi, Keito and Cao, Qixuan and Selcuk, Kemal and Hu, Tianrui and Niazi, Shaila and Aadit, Navid Anjum and Kanai, Shun and Ohno, Hideo and Fukami, Shunsuke and Camsari, Kerem Yunus},
  year = {2024},
  month = {Mar.} 
}

@article{600TMR_1,
  title = {High Performance {MgO}-barrier Magnetic Tunnel Junctions for Flexible and Wearable Spintronic Applications},
  author = {Chen, Jun-Yang and Lau, Yong-Chang and Coey, JMD and Li, Mo and Wang, Jian-Ping},
  journal = {Scientific Reports},
  month = {Feb.},
  year = {2017},
  volume = {7},
  number = {1},
  pages = {42001},
  issn = {2045-2322},
  doi = {10.1038/srep42001},
}

@article{600TMR_2,
    author = {Scheike, Thomas and Wen, Zhenchao and Sukegawa, Hiroaki and Mitani, Seiji},
    title = "{631 $\%$ room temperature tunnel magnetoresistance with large oscillation effect in CoFe/MgO/CoFe(001) junctions}",
    journal = {Applied Physics Letters},
    volume = {122},
    number = {11},
    year = {2023},
    month = {Mar.},
    pages={112404},
    issn = {0003-6951},
    doi = {10.1063/5.0145873},
}

@INPROCEEDINGS{Saleh_p-bit,
  author={Bunaiyan, Saleh and Al-Dirini, Feras},
  booktitle={IEEE Nanotechnology Materials and Devices Conference (NMDC)}, 
  title={{MTJ}-Based p-Bit Designs for Enhanced Tunability}, 
  year={2022},
  volume={},
  number={},
  pages={14-16},
  month = {Nov.},
  doi={10.1109/NMDC46933.2022.10052369}}

@INPROCEEDINGS{Saleh_p-sensing,
  author={Bunaiyan, Saleh and Al-Dirini, Feras},
  booktitle={IEEE Electron Devices Technology \& Manufacturing Conference (EDTM)}, 
  title={Probabilistic Autonomous Data Acquisition Using Stochastic {MTJ} Based p-Bits}, 
  year={2024},
  volume={},
  number={},
  month = {Mar.},
  pages={1-3},
  doi={10.1109/EDTM58488.2024.10512188}}

@article{Neuromorphic_Spintronics,
  author = {Julie Grollier and Damien Querlioz and Kerem Y. Camsari and Karin Everschor-Sitte and Shunsuke Fukami and Mark Stiles},
  Journal = {Nature Electronics},
  title = {Neuromorphic Spintronics},
  year = {2020},
  volume = {3},
  number = {7},
  month = {Mar.},
  pages={360-370},
  language = {en},
}

@ARTICLE{Mohammed_2025,
  author={Mohammed, Mahmood A. and Al-Dirini, Feras and Emara, Ahmed S. and Roberts, Gordon W.},
  journal={IEEE Transactions on Circuits and Systems I: Regular Papers}, 
  title={Design for Slew-Rate in Multi-Stage {CMOS OTAs}}, 
  year={2025},
  volume={},
  number={},
  pages={1-14},
  doi={10.1109/TCSI.2025.3593405}}

@patent{Patent_tunable_p-bit,
author= {Saleh Ahmad S Bunaiyan and Feras Mohamad Ameer Al-Dirini}, 
title= {Apparatus and method of implementing a probabilistic bit (p-bit) circuit with enhanced tunability}, 
language= {English}, 
assignee= {King Fahd University of Petroleum and Minerals}, 
address = {Dhahran, Saudi Arabia}, 
nationality = {U.S.}, 
type = {Patent Application}, 
number = {18 417 631}, 
day = {22}, 
dayfiled = {19}, 
month = {May}, 
monthfiled = {Jan.},
year = {2025}, 
yearfiled = {2024},
url = {https://patents.google.com/patent/US20250169370A1/en}
}
\end{document}